# Antiferromagnetic multi-level memristor using linear magnetoelectricity

Y. T. Chang[1#], J. F. Wang[1#], W. Wang[1#], C. B. Liu[2], B. You[1], M. F. Liu[3], S. H. Zheng[3,4], M. Y. Shi[1], C. L. Lu[1*], and J. –M. Liu [3,4]

[1] *School of Physics & Wuhan National High Magnetic Field Center, Huazhong University of Science and Technology, Wuhan 430074, China*

[2] *College of Physics and Electronic Engineering, Nanyang Normal University, Nanyang 473061, China*

[3] *Institute for Advanced Materials, Hubei Normal University, Huangshi 435001, China*

[4] *Laboratory of Solid State Microstructures, Nanjing University, Nanjing 210093, China*

[#]These authors contributed equally: Y.T. Chang, J.F. Wang, W. Wang.
[*] Email: cllu@hust.edu.cn

# Abstract

The explosive growth of artificial intelligence and data-intensive computing has brought crucial challenge to modern information science and technology, i.e. conceptually new devices with superior properties are urgently desired. Memristor is recognized as a very promising circuit element to tackle the barriers, because of its fascinating advantages in imitating neural network of human brain, and thus realizing in-memory computing. However, there exist two core and fundamental issues: energy efficiency and accuracy, owing to the electric current operation of traditional memristors. In the present work, we demonstrate a new type of memristor, i.e. charge $q$ and magnetic flux $\varphi$ space memristor, enabled by linear magnetoelectricity of $Co_4Nb_2O_9$. The memory states show distinctly linear magnetoelectric coefficients with a large ratio of about 10, ensuing exceptional accuracy of related devices. The present $q$-$\varphi$ type memristor can be manipulated by magnetic and electric fields without involving electric current, paving the way to develop ultralow-energy-consuming devices. In the meanwhile, it is worth to mention that $Co_4Nb_2O_9$ hosts an intrinsic compensated antiferromagnetic structure, which suggests interesting possibility of further integrating the unique merits of antiferromagnetic spintronics such as ultrahigh density and ultrafast switching. Linear magnetoelectricity is proposed to essential to the $q$-$\varphi$ type memristor, which would be accessible in a broad class of multiferroics and other magnetoelectric materials such as topological insulators. Our findings could therefore advance memristors towards new levels of functionality.

**Introduction**

The terminology "memristor" was first proposed as the fourth circuit element by Chua at a half century ago *(1)*, and was experimentally confirmed as pinched *i-v* hysteresis loops in a simple oxide TiO$_2$-based two-terminal device by Strukov *et al.* decades later *(2)*, where *i/v* is the current/voltage across the memristor. Such memristive behavior has been widely observed in various compounds beyond the oxides, and typically featured with large resistance ratio (> 10) and relatively high switching speed (~ ns) *(3-6)*. These are interesting properties not only for mimicking the neural network and therefore developing in-memory computing, which supports in-situ calculations in the memory unit and thus removes time and energy dissipation due to data movement *(7-11)*. While memristor has attracted abundant attention from both fundamental research and industry, there exist some core issues that must be addressed for viable application.

A prime difficulty to improve the energy efficiency is that the memory states are commonly manipulated using electric current *i*, and especially large *i* is required when the resistance is low *(7)*. Consequently, the Joule heating becomes a major reason for device integration and conductance nonlinearity *(12, 13)*, which is a serious problem for the accuracy and high-performance of memristive devices. For instance, identical input values would yield different conductance response because of the nonlinear *i-v* curves. To solve these issues, a *q-φ* analysis method, where *q* is the electric charge (equivalently open-circuit voltage) and *φ* is the magnetic flux, was recently proposed to apply to memristor network characterized with various types of *i-v* loops, and considerably reduced power consumption was achieved *(14)*. This recalls the initial motivation for the concept of memristor: to describe the relation between two fundamental circuit variables *q* and *φ* *(1)*. Indeed, the connection of these two quantities can be realized via some roadmaps other than the current effect.

Interestingly, rather than reengineering the artificial neural network composed of *i-v* type memristors, there is an alternative route to link *q* and *φ* directly by means of advanced magnetoelectric (ME) coupling as an intrinsic property of multiferroic materials *(15, 16)*. This alternative roadmap presents at least three major advantages that offer the proposed single-phase multiferroic based *q-φ* memristor superior performances and are either inaccessible if it is based on composite magnetoelectrics such as ferromagnetic / ferroelectric multilayers *(17)*. First, it is allowed to realize fully electric field-control of the ferroic-orders such as magnetization (*M*) and electric polarization (*P*), which is pivotal for developing devices with ultralow power consumption. Second, the ME susceptibility allows a readout by electric voltage (*17, 18*). Third, the majority of multiferroics are antiferromagnetic (AFM), and the magnetic texture can be either collinear or noncollinear. This may further permit to accommodate the superior properties such as ultrahigh density and ultrafast switching due to the inherent zero stray field and THz spin dynamics of antiferromagnets (*19-24*), which is essentially important for the in-memory computing. However, an AFM memristor enabled by ME coupling of multiferroics remains unexplored up to date, even in the conceptual level.

Following this line, such a memristor device would be perfect if further a linear *q*-*φ* dependence can be accessed. Indeed, linear magnetoelectricty $\alpha \sim dP/dH$, which can be equally expressed as $\sim dq/d\varphi$ after simply dividing the permeability of free space, is favored to achieve remarkable accuracy. In the meantime, compensated AFM phase with zero net magnetization is more suitable for exploiting the merit of antiferromagnets. In this study, we demonstrate a *q*-*φ* space memristor enabled by linear ME coupling of a honeycomb antiferromagnet $Co_4Nb_2O_9$, while this linear ME response can be controlled by both magnetic field *H* and electric field *E*, with ultralow power consumption. In the following, we first recall some fundamental aspects of $Co_4Nb_2O_9$ relevant to this motivation, and then present the key experimental results that demonstrate the AFM memristor, its control by *H* and *E*, and interesting memristive behaviors.

## Results

*Magnetic structure and ME response*

$Co_4Nb_2O_9$ crystallizes in a trigonal symmetry, and $Co^{2+}$ cations occupy two inequivalent sites, i.e. Co(1) and Co(2), as shown in Fig. 1(a) (*25-27*). The Co(1)$O_6$ octahedra are corner-linked while the Co(2)$O_6$ cages are edge-shared, forming two different honeycomb layers alternatively stacked along the *c*-axis. It is clear that the $Co^{2+}$ ions are antiferromagnetically coupled within the basal planes, and the AFM phase follows the magnetic symmetry *C*2/*c'* and takes an easy-plane anisotropy (*27-29*). The AFM transition is featured by a cusp-like peak in the *M*(*T*) curve at $T_N \sim 27$ K with *H*//*ab*-plane. Details of the magnetic behaviors are shown in the Supplementary Information.

While *M* ~ 0 is evidenced in the ground state, the typical AFM characters are illustrated by the linear and non-hysteretic *M(H)* curves with *H*//[110] up to the highest field $H_0 \sim 52$ T at *T* = 4.2 K

(Fig. 1(b)). A weak anomaly at $H \sim 0.4$ T, marking the spin flop, can be identified (see Fig. S1) *(29)*. The AFM state is robust and the magnetic symmetry remains unchanged with $H$ up to $\sim 10$ T *(27)*. The magnetic structures, drawn in Fig. 1(a) (bottom schematic), include the AFM-1 state where the Co(1) and Co(2) spins show a canting angle of $\sim 10.5°$ from each other. The $M(H)$ data, shown in Fig. 1(b), find two additional anomalies at $H_{c1} \sim 23$ T and $H_{c2} \sim 41$ T. A kink feature in the $dM/dH$ curve at $H \sim H_{c1}$ marks the spin-flip from the AFM-1 state to the AFM-2 state, by likely a collective switching of the six $Co^{2+}$ spins as a ring. As a consequence, the six spins are all canted toward $H//[110]$, resulting in the slight $M(H)$ slope increase at $H_{c1} \sim 23$ T. This field-driven spin-canted state is a necessary ingredient for the ME memristive effect. The transition at $H_{c2}$ is sharp with $M \sim 12.3$ $\mu_B/f.u$ matching well with previous neutron data *(27-29)*, suggesting the AFM - FM transition where the six spins are totally switched to the FM alignment along [110].

It is mentioned that the AFM-1 (AFM-2) state has various variants and one of them is drawn as AFM-3 (AFM-4) in Fig. 1(a). In the whole $H$-range below $H_{c2}$, the coexistence of various AFM domain variants is expected while these domains are merged, leading to the FM phase as $H > H_{c2} \sim 41$ T in the $H$-increasing sequence. The situation could be different for the $H$-decreasing sequence that a single AFM domain state is more favored because this AFM state has nonzero canted moment and thus is favored, as evidenced by the magnetoelastic strain data shown in Fig. 1(c) that reaches up to $\Delta L/L \sim 180$ ppm in the high-$H$ range, where $L$ and $\Delta L$ are the sample's length and its variation. At $H \sim H_{c2} \sim 41$ T, $\Delta L/L$ turns to decrease with $H$, implying the lattice contraction. Such lattice modulation would be essential for the development of domain walls. As revealed by Khanh et al. *(30)*, the magnetoelastic effect of $Co_4Nb_2O_9$ is associated with the relative angle between spin and $H$, explaining the absence of anomaly at $H_{c1} \sim 23$ T in the $\Delta L/L(H)$ curve, since the spins of both AFM-1 and AFM-2 phases have the same relative angle to $H$.

More amazing is the emergence of $H$-driven nonzero polarization $P$ over the broad $H$-range, evidencing the linear ME effect expressed by $P = \alpha H$. The measured magneto-current $I$ by varying $H$ // [110] is plotted in Fig. 1(d) and the $P(H)$ data in Fig. 1(e). The magnetically driven ferroelectricity, based on the symmetry argument *(31, 32)*, has its major $P$ component along the [110] direction. The $I(H)$ data demonstrate distinctly different behaviors in various $H$-ranges, as shown in Fig. 1(d). First, the $I(H)$ appears to be a non-zero constant in the low-$H$ case ($H < H_{c1}$) except the very small field window, the consequence of the linear ME effect. Second, the constant

$I$-values respectively in the $H$-increasing and $H$-decreasing sequences, as indicated by the arrows imply the different linear $P(H)$ dependences and thus different $\alpha$-values in the two sequences. Third, the two anomalies in the high-$H$ range are identified in connection with the $M(H)$ data, evidencing that the ME effect is intrinsic.

Consequently, the evaluated $P(H)$ curves in the whole $H$-cycling are plotted in Fig. 1(e). The most striking feature is the remarked $P(H)$ hysteresis with two distinct linear ME branches, i.e. the so-called ME memristive behavior addressed in this work. This effect, taking the $q$ - $\varphi$ characteristic, is fundamentally different from the $i$-$v$ type memristive effect. The ME coefficient $\alpha = dP/dH$ for the two linear ME states in the $H$-increasing and $H$-decreasing sequences (paths) are ~ 4 ps/m and ~ 40 ps/m, respectively. Such a two-state behavior, with an $\alpha$-ratio of ~ 10, evidences a novel memristor functionality that offers specific advantages and application promising. Obviously, the two distinct ME states are due to the drastically enhanced $P$ event as $H > H_{c1}$, reaching the maximal at $H = H_{max}$ ~ 35 T in the $H$-increasing sequence. The big difference in the two sequences originates from the AFM domain reversal and thus the ferroelectric domain reversal via the domain wall motion scenario that is certainly a new version of mechanism proposed for memristor effect in the very beginning era (*1*). We add in Fig. 1(e) the schematic ferroelectric domains at different stages of the $H$-increasing/decreasing sequences, with the olive arrows indicating the two domains. In the $H$-increasing sequence, the initial random domains are gradually aligned by $H$ in the AFM-1 domain reversal into the AFM-2 domains, and polarization $P$ increases linearly until the fully aligned AFM-2 domains (also the fully aligned ferroelectric domains). In the $H$-decreasing sequence, the fully aligned domains are gradually de-aligned, and polarization $P$ deceases linearly. The two paths are involved with the domain reversal processes which are basically of the first-order transition nature, and therefore remarked ME hysteresis and ME memory become inevitable. It is conclusive that the true $q$ - $\varphi$ type AFM memristor, is driven by multiferroic domain evolution accompanied by the $P$-reversal process.

*Electro-control of ME memristive state*

As a ME memristor, one of its advantages is the possible electro-control of memristive state including the ME response and the AFM domain structure. To demonstrate this advantage, the sample was magnetically zero-field cooled down to $T$ = 4.2 K, and the ME effect was measured under a given electric field $E = \pm 10$ kV/cm. The obtained $I(H)$ curves are shown in Fig. 2 (a) in the $H$-cycle respectively. The measured two sets of $I(H)$ curves show the roughly opposite behaviors with the similar qualitative features such as two anomalies at $H_{c1}$ and $H_{c2}$. The $E$-controlled $I(H)$ sign reversal is clearly demonstrated. Subsequently, the evaluated $P(H)$ curves are symmetric with the $P = 0$ axis and the two sets of ME coefficient $\alpha(H)$ show the opposite signs, as shown in Fig. 2(b). In such sense, the dual controllability of the AFM memristor by both $H$ and $E$ is revealed.

*Switching between the two ME states*

A key ingredient of the memristor is the nonvolatile and inter-switchable memory states, as illustrated by the specifically designed ME switching experiments, shown in Fig. 3 as an examples. The initial state was obtained simply by $H = 0$ cooling of the sample from high-$T$ state (far above $T_N$) down to $T = 4.2$ K at which the measurement was performed. The [110]-oriented $H$ with magnitude $H_0$ was applied to the sample in the $H$-increasing/decreasing cycling as a writing (switching) process, while the resultant state is read out by a small $H$ ($H_0 \ll H_{c1}$) subsequently. For a better view, only typical data are shown here (the full data in the Supplementary Information), and some $P(H)$ curves are shifted vertically.

(1) For a writing with $H_0 < H_{c1} \sim 23$ T (e.g., cycle 2) from the initial state, the measured $P(H)$ data always follow the L1 trace and exhibit no remarkable hysteresis, giving rise to a small $\alpha \sim 4$ ps/m, i.e. the sample is in the low ME state.

(2) For a writing with $H_0 > H_{c2}$ (e.g., cycle 8 & cycle 9), no matter how the previous state is, the resultant state after the $H$-cycling is the low ME state with $\alpha \sim 4$ ps/m, i.e. the sample is written into the low ME state. Nevertheless, the measured $P(H)$ hysteresis depends on the previous ME state. For cycle 8, the $P(H)$ hysteresis shows no loop since the previous state is in the high ME state ($\alpha \sim 40$ ps/m). For cycle 9, the $P(H)$ hysteresis shows large loop since the previous state is in the low ME state ($\alpha \sim 4$ ps/m) after the cycle 8.

(3) For a writing with $H_0 \sim H_{max} = 35$ T at 4.2 K (e.g., cycle 4), no matter how the previous state is, the resultant state after the $H$-cycling is the high ME state with $\alpha \sim 40$ ps/m, i.e. the sample is written into the high ME state.

*Training effects*

The aforementioned low/high ME state writing sequences are all deterministic and basically. These sequences don't reflect the full core property of a memristor, i.e. history-dependent training effect. In fact, this core effect occurs in the $H_{c1} < H_0 < H_{c2}$ region. For convenience, we start again from different initial state, and our observations can be highlighted as the following:

(1) For cycling with $H_{c1} < H_0 < H_{max}$, the training sequences starting from a low ME state are shown in Fig. 4(a) at $H_0 = 30$ T. This effect is characterized by the gradually shrinking $P(H)$ loop upon the sequential $H$-cycling, accompanied with the gradual increasing of the ME coefficient from $\alpha \sim 4$ ps/m to $\alpha \sim 40$ ps/m, as demonstrated by the dependence of $\alpha$ on the cycling number (Fig. 4(c)). Certainly, this number for completing the training process depends on $H_0$, and the more the cycling number the closer $H_0$ is to $H_{c1}$. It is expected that this number would be an exponential function of $(H_0 - H_{c1})^{-1}$, typical for behaviors around a critical point.

(2) For cycling with $H_{max} < H_0 < H_{c2}$, the training behavior is somehow different, and the results are shown in Fig. 4(b) for $H_0 = 39$ T as an example. Here one sees switching from the high ME state to an intermediate ME state with reduced $\alpha \sim 13$ ps/m, accompanied by enlarged $P(H)$ loop. This ME state behaves quite stable against consecutive cycling with $H_0 = 39$ T (more data in the Supplementary Information), different from the case with $H_0 = 30$ T. Such difference would be expectable since $H_0 = 39$ T is a lot larger than $H_{max}$, large enough to re-write the ferroelectric and AFM state evidently. The training effect becomes significant as $H_0$ is closer to $H_{c2}$, while it becomes extremely slow if $H_0 \sim H_{max}$ where the system is always in the high ME state.

**Discussion**

Our experiment represents a conceptual proof of a new type of AFM $q$-$\varphi$ memristor which is robust, nonvolatile, and controllable using $H$ and $E$ instead of current $I$. The memristor state is determined by the linear magnetoelectricity that explicitly describes the relation between the two fundamental variables $q$ and $\varphi$. This concept is basically different from conventional $i$-$v$ type memristor (2, 4). Certainly, this behavior relies on the multiferroic domain motion manipulated by $H$ and $E$, different from the operation of $i$-$v$ type memristor, for which electrical current is the primary parameter to write and read the memory states. Therefore, the present AFM memristor within the $q$-$\varphi$ space is promising to reduce the power consumption drastically with sufficiently high precision. On the other hand, the remarkable training behavior of this memristor suggests a connection with brain-like devices. The memristor possesses the linear $q$-$\varphi$ ($P$-$H$) relation over a broad $H$ range, crucial for accuracy and thereby efficiency of related devices such as artificial neural networks, one of the chief targets for memristor activities nowadays (*12, 13*). Such $q$-$\varphi$ memristor would be accessible in many other linear magnetoelectrics, such as weak ferromagnet $Ni_3B_7O_{13}I$ (*33*) and ferrimagnet $(Fe,Zn,Mn)_2Mo_3O_8$ (*34, 35*).

The inherent AFM structure of the present memristor may further allow to incorporate some superior properties due to the zero stray field and THz spin dynamics of antiferromagnets. AFM spintronics is an emergent field that aims to develop devices with exceptional memory storage and ultrafast switching (*20, 36-38*). As given in the 1970 Nobel lecture of Louis Néel, effects depending on the square of the spontaneous magnetization should thus show the same variation in antiferromagnets as in ferromagnetic substances (*39*). Inspired by this key principle, a large array of interesting effects such as large anisotropic magnetoresistance and anomalous Hall effect have so far been revealed in antiferomagnets (*19, 21, 24, 40, 41*), suggesting that AFM materials can play an active role in spintronics devices as ferromagnets. Importantly, the antiparallel alignment of spins in AFM materials promises ultrahigh rigidity and density of related devices. Because of the interacted magnetic sublattices, the intrinsic frequency of AFM dynamics is about three orders of magnitude higher than ferromagnets (~GHz), which would shorten the operation time of AFM spintroncis devices drastically.

At last, it should be pointed out that the memristive behaviors demonstrated in the present work are essentially determined by the ME susceptibility, which can be read out by electrical voltage (*18*). This is fundamentally different from conventional multiferroic memory effects which rely on remnant electric polarization, akin to classic ferroelectric memory (*16*). For instance, in $YMnO_3$ thin films, memory states were realized by controlling resultant cycloidal domains (*42*). Moreover, such multiferroic memory associated with remnant polarization usually originates from higher order ME coupling, which is necessary but not sufficient for obtaining memristive behaviors (*43*).

In summary, multilevel ME memory functionality has been realized in honeycomb antiferromagnet $Co_4Nb_2O_9$, which show high rigidity against external magnetic perturbations. The memory states are nonvolatile and possess very different linear ME coefficients for the high and low ME states, respectively. By applying electric field, the memory states can be switched safely, and the sign of ME coefficient is changed simultaneously. The AFM memristive behaviors can be ascribed to the multiferroic domain evolution related to the inequivalent honeycomb lattices of Co ions. The present AFM memristor is appealing due to the possibility of designing low-power, fast-switching, and high density deices, which is highly desired for in-memory computing and thus the artificial intelligence. Therefore, our work unveils a new area to fully exploit the merit of memristors.

**Materials and Methods**

$Co_4Nb_2O_9$ single crystals were grown by an optical floating zone technique. The crystals were oriented using X-ray Laue photographs, and cut into thin plates with typical dimensions ~2.3×1×0.2 $mm^3$. Before the high field characterizations, some basic measurements such as magnetization as a function of $T$ and $H$ were carried out with $H$ applied along various crystalline directions using a superconducting quantum interference device (Magnetic Property Measurement System, Quantum Design).

High magnetic field measurements for magnetization were performed at various temperatures by means of a standard inductive method employing a couple of coaxial pickup coils. The ferroelectric polarization was measured using the pyroelectric technique. Gold electrodes were sputtered onto the widest faces of the samples. The sample was first cooled from high temperature (far above $T_N$~27 K) down to target temperature without applying magnetic field, and then the $I(H)$ measurements at different temperatures were carried out. During the electric measurements, a bias field $\pm E$=10 kV/cm was applied. The interval between two high field measurements is roughly 1-2 hours. The pulsed high magnetic field up to ~52 T with a duration time of ~10 ms was generated by a nondestructive short-pulse magnet in the Wuhan National High Magnetic Field Center (WHMFC). Magnetostriction effect ($\Delta L/L$) was measured up to ~50 T along the [110] direction using an optical fiber grating technique. During the magnetostriction measurements, the crystal was attached to a platform for mechanical stabilization during the rapid magnetic field pulse.

**Supplementary materials**

Supplementary material for this article is available at:

Fig. S1. Magnetization data measured with *H* applied along various directions.

Fig. S2. High field magnetization data along various directions.

Fig. S3. Switching from a low-$\alpha$ state to a high-$\alpha$ state.

Fig. S4. Switching from a high-$\alpha$ state to a low-$\alpha$ state.

Fig. S5. Realization of an intermediate ME state.

Fig. S6. *P*(*H*) curves measured at *T*=4.2 K with *H*=33 T applied along the [110] direction.

Fig. S7. *P*(*H*) curves obtained by continuous *H*-cycling at *T*=4.2 K with *H*=30 T.

Fig. S8. *P*(*H*) curves obtained by continuous *H*-cycling at *T*=4.2 K with *H*=39 T.

Fig. S9. A complete set of *M*(*H*) and *P*(*H*) data measured at various temperatures, and a summarized phase diagram.

**Competing interests:** The authors declare no competing interests.

**Acknowledgements:** This work is supported by the National Nature Science Foundation of China (Grant Nos. 11774106, 11874031, 11834002, and 51721001), the National Key Research Projects of China (Grant No. 2016YFA0300101), Hubei Province Natural Science Foundation of China (Grant No. 2020CFA083), and the Fundamental Research Funds for the Central Universities (Grant No. 2019kfyRCPY081, 2019kfyXKJC010).

**Author contributions:** C.L.L. conceived the project. M.F.L. and S.H.Z. prepared the samples. Y.T.C., J.F.W., W.W., C.B.L., and M.Y.S. performed the physical property measurements. B.Y., M.F.L., and S.H.Z. contributed to the data analysis. C.L.L. and J.M.L. wrote the manuscript.

**Figure Captions**

**Fig.1. Characterization of physical properties of multiferroic $Co_4Nb_2O_9$.** (a) Crystalline structure of $Co_4Nb_2O_9$, where the two inequivalent Co sites are labeled. Two different honeycombs with Co(1) and Co(2) are schematically shown: the Co(1) hexagon is corner linked and the Co(2) ring is edge shared. Sketched antiferromagnetic honeycombs AFM-1 are shown according to previous neutron scattering experiments. AFM-3 shows possible alignments equally existing in the material. The next-nearest neighboring Co ions (indicated by red and blue arrows) are ferromagnetically coupled along the *c*-axis, and exhibit a relative canting angle of ~ 10.5° at $H = 10$ T. Driven by $H > H_{c1}$, AFM-2 phase is stabilized, which has other variants such as AFM-4. At $H > H_{c2}$, AFM to FM transition is triggered, and thus a paraelectric phase arises. (b) Magnetization as a function of $H$ measured at $T = 4.2$ K. The corresponding $dM/dH$ data are also shown, and two anomalies can be identified at $H_{c1} \sim 23$ T and $H_{c2} \sim 41$ T. (c) Magnetostriction $\Delta L/L$ as a function of $H$. (d) Pyroelectric current $I$ as a function of $H$. (e) Integrated ferroelectric polarization as a function of $H$. The arrangement of *P*-components is also sketched (olive arrows). All the properties were measured with $H$ applied along the [110] direction at $T = 4.2$ K.

**Fig. 2. Electric field control of the magnetoelectric effect.** (a) Pyroelectric current $I$ as a function of $H$ measured with different electric fields $\pm E = 10$ kV/cm at $T = 4.2$ K. (b) The corresponding integrated $P(H)$ curves are shown, which can be completely reversed by electric field. The ME coefficient $\alpha$ are also labeled with the $P(H)$ curves.

**Fig. 3. Switching of the antiferromagnetic memristor.** A series of $P(H)$ curves obtained with different *H*-cycles at $T=4.2$ K. The measurements were performed continuously. For a better view, only typical $P(H)$ curves are shown, and some of them are shifted vertically. (a) Low to high-$\alpha$ setting. (b) High to low-$\alpha$ setting. (c) Intermediate-$\alpha$ setting. The blue and red dashed lines indicate the L2 and L1 traces, respectively. (d) Summarized memristive switching behaviors. The magnetic field for each *H*-cycle is shown in the bottom figure.

**Fig. 4. Training effect of the memristor.** (a) Starting from the low-$\alpha$ state, successive cycles with $H_0=30$ T were carried out, and a series of $P(H)$ curves with gradually shrunk hysteresis were recorded. (b) $P(H)$ curves obtained by continuous cycles with $H_0$=39 T. (c) Magnetoelectric coefficient $\alpha$ as a function of the $H$-cycling number.

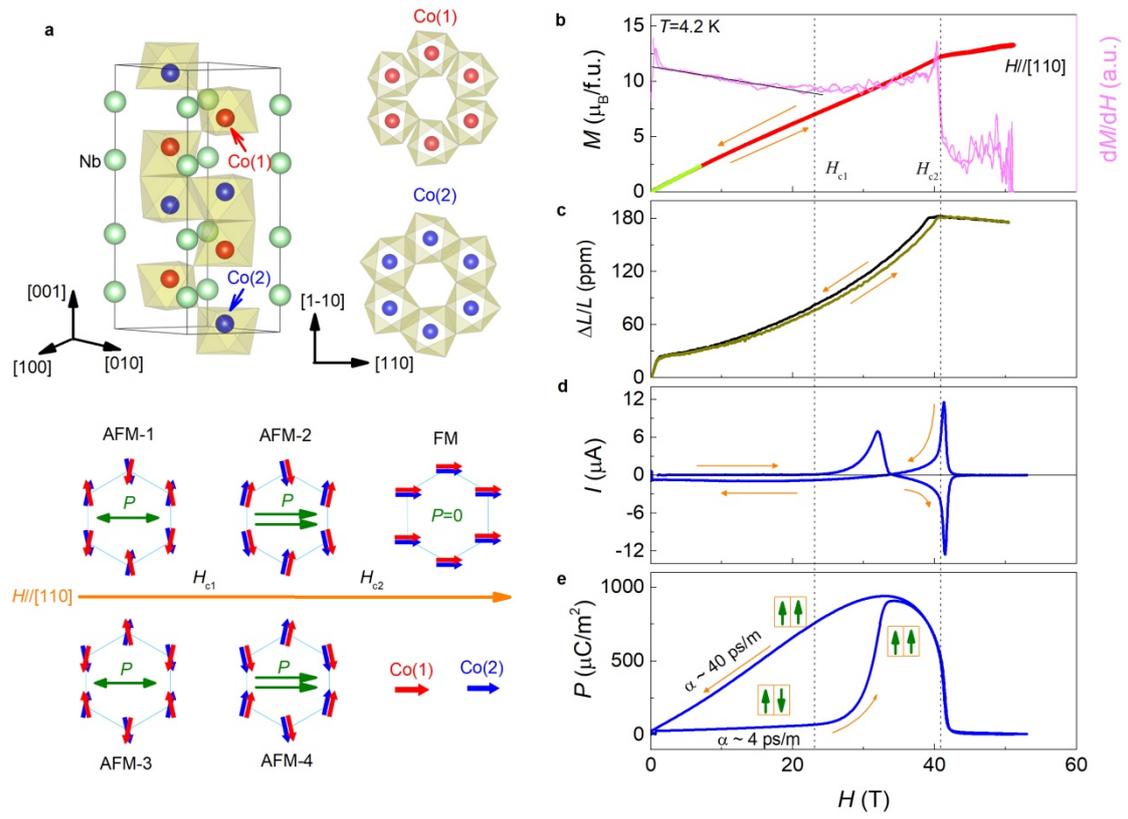

Figure1

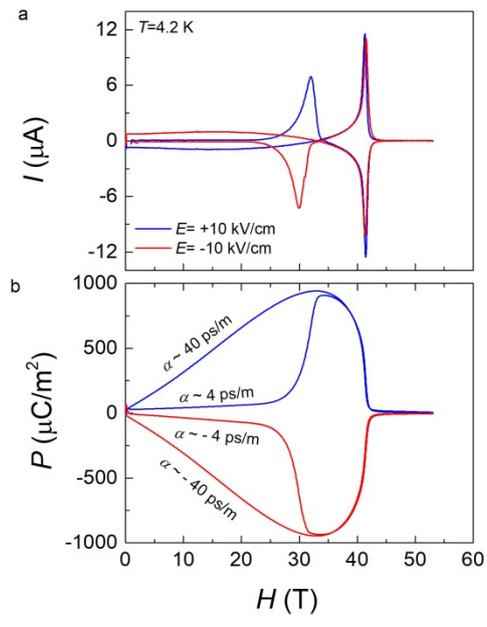

Figure 2

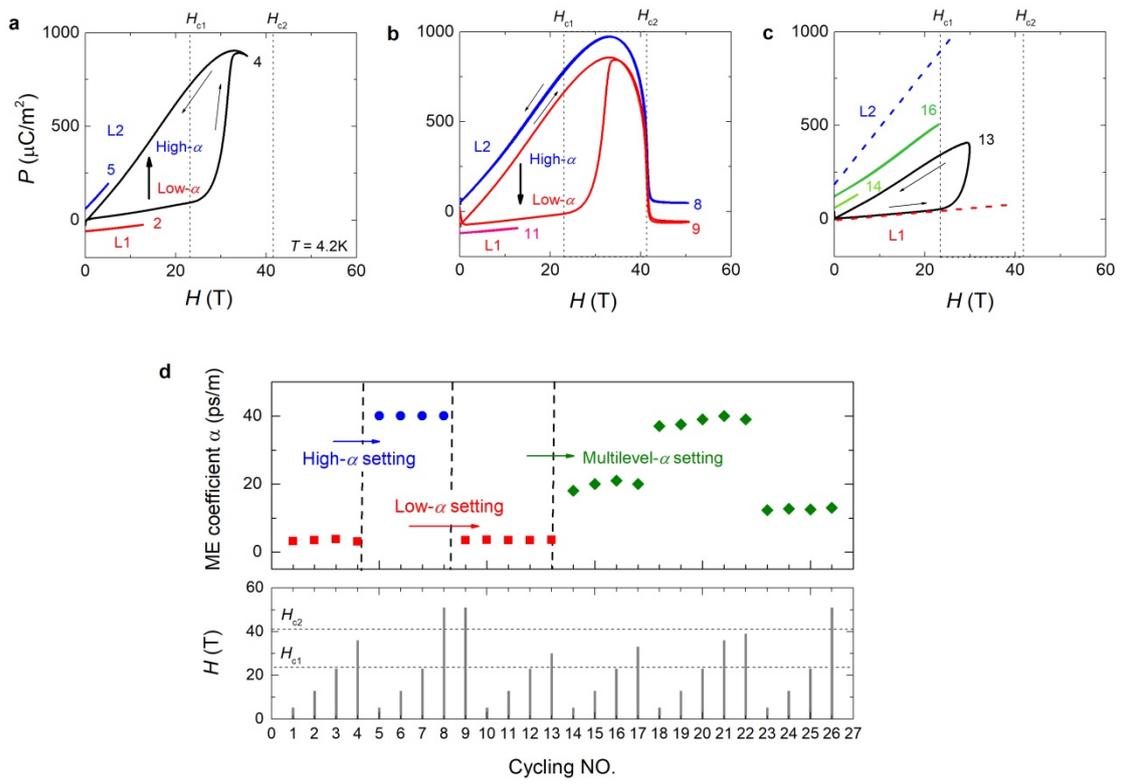

Figure 3

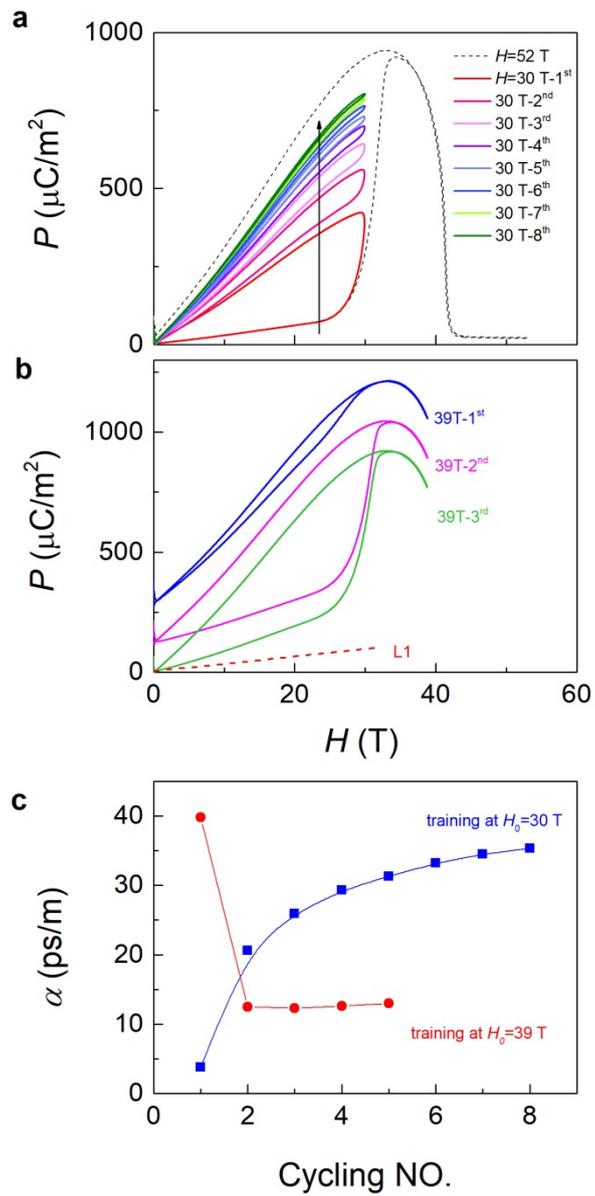

Figure 4